# Electrostatic gating of metallic and insulating phases in SmNiO$_3$ ultrathin films

Sieu D. Ha[*,1], Ulrich Vetter[2], Jian Shi[1], and Shriram Ramanathan[1]

[1]School of Engineering and Applied Sciences, Harvard University, Cambridge, MA 02138, USA

[2]Physikalisches Institut, Georg-August-Universität Göttingen, Friedrich-Hund-Platz 1, Göttingen D-37077, Germany

**Abstract**

The correlated electron system SmNiO$_3$ exhibits a metal-insulator phase transition at 130 °C. Using an ionic liquid as an electric double layer (EDL) gate on three-terminal ultrathin SmNiO$_3$ devices, we investigate gate control of the channel resistance and transition temperature. Resistance reduction is observed across *both* insulating and metallic phases with ~25% modulation at room temperature. We show that resistance modulation is predominantly due to electrostatic charge accumulation and not electrochemical doping by control experiments in inert and air environments. We model the resistance behavior and estimate the accumulated sheet density (~1-2 x 10$^{14}$ cm$^{-2}$) and EDL capacitance (~12 µF/cm$^2$).

---

[*] Corresponding author: sdha@seas.harvard.edu

There is extensive interest in understanding the complex electronic phase diagrams of correlated oxides and their interfaces towards potential applications for next-generation electronics, photonics, and energy. Among correlated oxide materials, the rare-earth nickelates ($R$NiO$_3$, $R$ = rare-earth element) have a unique phase diagram with an antiferromagnetic insulating ground state and a paramagnetic metallic high temperature phase.[1] The metal-insulator phase transition temperature ($T_{MI}$) varies monotonically from ~130-600 K between PrNiO$_3$ and LuNiO$_3$ and the resistivity change can be several orders of magnitude across the transition. SmNiO$_3$ (SNO) is the first in the series with $T_{MI}$ above room temperature (~130 °C or ~400 K) and is a candidate for integration of correlated phenomena into conventional room temperature electronics. Electric-field or electrostatic control of the metal-insulator transition (MIT) in SNO will be useful for examining fundamental physical properties in the high carrier density limit and perhaps for future device technologies. In this work, we present a study on electrostatic modulation of the phase transition properties in SNO ultrathin films above room temperature using an ionic liquid electric double layer gate capacitor.

Ionic liquid (IL) gates can be used to achieve extremely high carrier densities ($10^{14-15}$ cm$^{-2}$) in three-terminal structures. Ionic liquids are likely necessary for gating of the nickelates, which have carrier densities on the order of $10^{21-22}$ cm$^{-3}$ (~$10^{14}$ cm$^{-2}$).[2,3] Room temperature ILs are small molecule organic salts in which, in the liquid phase, the cation and anion molecules are mobile and can be used to attract or store charge at IL-electrode interfaces.[4] The pairing of molecular ion and mobile charge is referred to as an electric double layer (EDL), and the effective capacitance of an EDL can be large, typically greater than 10 μF/cm$^2$.[5] However, care must be taken to avoid chemical or electrochemical reactions, which may arise due to water contamination in the IL or at sufficiently high temperatures due to the inherent Faradaic nature of ILs.[6-9]

One method to avoid unwanted electrochemistry is to utilize an IL that is predominantly solid in the required temperature range for the material in question.[2] In this case, the gate voltage is first applied at elevated temperature above the melting point of the IL. Then the temperature is lowered to the desired regime below the IL melting point and the ions are no longer mobile, maintaining the electrostatic effect but significantly reducing reactivity. We utilize this methodology for our SNO IL-gated structures here. We also examine the effects of intentional defect generation in SNO films to demonstrate that defect modulation is not occurring in our IL-gating experiments.

Complete experimental details can be found in the Supplementary Materials.[10] $SmNiO_3$ thin films were grown by RF magnetron sputtering in relatively high background pressures onto single crystal $LaAlO_3$ (001) (LAO).[11] Thickness was determined by X-ray reflectivity. Due to high carrier concentration in the nickelates, electric field screening lengths are small (few Å) and ultrathin films are required to observe resistance modulation from an electrostatic effect.[12] Films used for IL gating were $3.3 \pm 0.1$ nm (8-9 pseudocubic unit cells) thick. A high angle annular dark field scanning transmission electron microscope (STEM) image of the film is presented in Fig. 1a, showing epitaxial growth of SNO on LAO and a clean, abrupt interface. Films used to demonstrate the effects of defects were 16-21 nm thick. Ion irradiation of the SNO films was performed using 6.8 MeV Cl ions with a charge state of 2+ and ion flux of $\sim 10^{14}$ ions/cm$^2$·s. Three-terminal structures for IL gating were patterned by standard photolithography, as illustrated in Fig. 1b. The SNO patterned bars had dimensions of 400 μm x 2000 μm. The IL used in this work (Covalent Associates, Inc.) is composed of the anion tris(trifluoromethylsulfonyl)methide and the cation 1,2,3,4,5-pentamethylimidazolium, as shown in Fig. 1c (drawn with ChemBioDraw), with a melting point of ~120 °C. Resistance-temperature ($R$-$T$) measurements were performed 1) in air with the IL used as-received and 2) in 200 sccm $N_2$ gas flow after baking the IL in $N_2$ at 130 °C for 12–14 h to remove residual parasitic water from the synthesis procedure. Significantly different results were observed for the two cases, as discussed below.

Crystalline IL powder was initially placed directly on top of the SNO channel at room temperature. For each gate voltage increment, $V_G$ was applied (with source terminal grounded) above the IL melting point for 30–40 min before initiating the $R$-$T$ measurement, which is sufficient time to fully charge the EDL capacitance.[13] For measurements in $N_2$ flow, the gating procedure and subsequent temperature cycling were performed without breaking $N_2$ purge. Here, we discuss results only for negative $V_G$, which corresponds to anion accumulation on the SNO surface and hole accumulation in the channel. Limited studies with positive $V_G$ resulted in irreversible degradation of the SNO bar, likely due to electrochemical reactions, which was also observed in IL-gated $NdNiO_3$ experiments.[2]

As a frame of reference for considering defect effects in IL experiments, we show $\rho$-$T$ of SNO as a function of controlled defect concentration in Fig. 2. As we have previously shown, SNO films sputtered in higher background pressures have qualitatively higher oxygen content.[11] The black curve in Fig. 2 is from an SNO film sputtered at ~370 mTorr, and the resistivity ratio from room temperature to the metallic phase is ~10. As growth pressure decreases to ~285 mTorr (blue curve), oxygen concentration decreases and the resistivity ratio drops to ~4. Note that the resistivity decreases in the insulating phase but increases in the metallic phase. This is likely due to oxygen vacancies behaving as shallow dopants in the insulating phase and scattering sites in the metallic phase. Quite similar behavior is observed for films grown at ~370 mTorr background pressure and irradiated with $Cl^{2+}$ ions (red curve). This is consistent with the fact that the mainly-electronic energy loss due to the chlorine ions (see Supplementary Material)[10] will locally heat the sample around the ion path allowing oxygen to escape from the sample, leaving oxygen vacancies. With respect to IL-gating measurements, an electrostatic gating effect should modulate the resistance in the same direction in both insulating and metallic phases, unlike the asymmetric modulation observed here with the introduction of defects.

We first perform experiments with the IL used as-received. Sheet resistance ($R_s$) as a function of temperature and gate voltage is shown in Fig. 3a for this case. At $V_G = 0$ V, $R_s$ changes by a factor of ~3 from room temperature to the metallic phase. The relatively low change in resistance is due to the enhanced effects of epitaxial strain in films of low thickness and has been previously reported for NdNiO$_3$ films of similar thickness.[2,14] As a function of increasing negative $V_G$ up to -0.8 V, the hysteresis window slightly widens and the overall resistance increases over the full temperature range. At $V_G = -1.2$ V, the SNO bar becomes an open circuit and the film visibly appears to be etched away. The overall increase in resistance contrasts previous work on IL-gated NdNiO$_3$ and Hall coefficient measurements on SNO thin films.[2,3] The IL-gated NdNiO$_3$ report shows a pronounced decrease in resistivity with increasing negative $V_G$. This is in agreement with the Hall measurements on SNO, which show that carriers in the temperature regime examined here are hole-like, and therefore increasing negative $V_G$ should increase hole carrier concentration in the channel and reduce resistance. The difference in $R_s$ vs. $V_G$ behavior with expected behavior from literature suggests that, for experiments in air with the IL used as-is, the observed resistance modulation is not due to electrostatic gating. This is further supported by the observed etching of the SNO bar at $V_G = -1.2$ V, indicative of electrochemical reactions. The increase in resistance for voltages lower than -1.2 V may therefore be due to incomplete etching of the device, which leads to lower thickness and higher sheet resistance for fixed resistivity.

Resistance-temperature measurements after baking the IL and performing measurements in a sealed probe station under $N_2$ purge are markedly different than with the IL used as-received, as seen in Fig. 3b. We show only curves on cooling for clarity and because the hysteresis does not change appreciably with gate voltage. As negative $V_G$ increases, there is a clear monotonic decrease in sheet resistance in both insulating and metallic phases. The conduction enhancement, unlike in the case of the IL used as-is, agrees well with previous experiments on low temperature IL-gating of $NdNiO_3$ and from what is expected from Hall measurements of SNO.[2,3] *A critical aspect of the data here is that the resistance is lowered in both insulating and metallic phases.* As shown in Fig. 2, defects can behave as dopants in the insulating phase, thereby reducing the resistance and misleadingly suggesting the existence of a gating effect. However, the same defects can behave as scattering sites that increase resistance in the metallic phase.[15,16] The reduction in $R_s$ at all temperatures with increasing negative $V_G$, as seen in Fig. 3b, is substantial evidence of room-temperature electrostatic gating of SNO. This is further supported by data in the Supplementary Material showing that the resistance modulation is reproducible, not due to gate leakage current, and that the gate current density is well below the threshold for electrochemical oxidation/reduction.[10]

The significant differences in electrical properties between using the IL as-received and after baking demonstrate the complexities of IL experiments and interpretation. ILs are electrochemically active,[7] and there can be residual reactive impurities from synthesis, water being the most common.[17] It was reported that the IL composed of 1-ethyl-3-methylimidazolium and bis(trifluoromethylsulfonyl)imide, which are somewhat chemically similar to the IL used here, are electrochemically stable from -2.0 V to +2.1 V for 20 µA/cm$^2$ current density.[4] Thus, the SNO degradation at $V_G$ = -1.2 V for the IL used as-received implies that the reactivity is caused by impurities aside from the nominal IL constituents. Water impurities are known to significantly decrease the stability window of ILs, likely due to electrolysis.[18] Moreover, hydroxide ions, which may form from water electrolysis, have been shown to degrade LaNiO$_3$ through reactions with Ni.[19] This suggests that the SNO degradation observed with the unbaked IL may be due to water contamination. Water absorption into the IL can also occur in ambient conditions,[18] thus underscoring the importance of both baking the IL and performing measurements in inert conditions, as noted in a study by Ji *et al*.[6]

The percent change in sheet resistance with respect to the $V_G$ = 0 V value (*i.e.* ($R_s(V) - R_s(0))/R_s(0)$) is shown in Fig. 4a as a function of temperature and gate voltage. It can be seen that the $R_s$ modulation is monotonic with respect to $V_G$ and temperature. The resistance modulation due to electrostatic gating is largest in the insulating phase, in agreement with IL-gated NdNiO$_3$ experiments.[2] This can be understood simply if the electrostatically accumulated charge layer is considered as a conduction path parallel to the bulk of the SNO bar. A parallel channel will reduce the total resistance of the bar, and the percent reduction will increase in magnitude as the 0 V resistance increases (temperature decreases). There may also be temperature- and phase-dependence of the EDL capacitance that affects the accumulated sheet carrier density as temperature decreases. The $T_{MI}$ modulation with gate voltage is shown in Fig. 4b. There is a monotonic decrease in $T_{MI}$ with increasing negative $V_G$ from ~132 °C at $V_G$ = 0 V to ~120 °C at $V_G$ = -2.5 V. The gate voltage dependence of $T_{MI}$ appears to be quadratic. Suppression of $T_{MI}$ with increasing $V_G$ indicates that electrostatic charge accumulation is driving the system further into the metallic phase. This suggests the possibility of electrostatic triggering of the metal-insulator transition.

We can model the $V_G$-dependence of the sheet resistance by considering that the hole density of the gated SNO bar has a gradient profile due to the electrostatic charge accumulation near the IL-SNO interface. We consider only holes because our previous Hall effect measurements have shown that the carriers are majority hole-like from the metallic phase to well below room temperature.[3] It has been suggested that the carrier density in a gated nickelate exponentially decays from the gate-film interface with a characteristic length of the inverse Thomas-Fermi screening wave vector ($k_{TF}$).[20] This length scale is more appropriate for the nickelates than the Debye screening length due to high carrier concentrations even in the insulating phase. Such a spatial dependence of the hole density is given by $p(z,V) = p'(z,V) + p_o = p'(0,V)e^{-k_{TF}z} + p_o$, where $p'(0,V)$ is the excess hole density at the IL-SNO interface and $p_o$ is the uniform hole density in the absence of electrostatic gating. This density profile is illustrated in the inset of Fig. 4c. For a carrier density gradient along the thickness of a bar, the sheet resistance is $R_s(V) = 1/e \int_0^t \mu(z)p(z,V)dz$ for a bar of thickness $t$ and mobility $\mu(z)$. For a degenerate Fermi gas, the Thomas-Fermi wave vector $k_{TF} = \sqrt{3e^2p/2\epsilon E_F}$. For typical values of the dielectric constant $\epsilon$ and Fermi energy $E_F$ in the nickelates,[21,22] and using $p$ as measured from our previous Hall effect measurements,[3] we find $k_{TF} \sim 1$ Å, of the same order as generally found in metallic or high density systems. The significance here is not the exact value of $k_{TF}$ but the fact that $k_{TF}^{-1} \ll t$, which allows for simplification of the sheet resistance expression. If we assume that the mobility is independent of position and excess hole density, we can solve for the excess sheet charge density $Q(V) = \int_0^t p'(z,V)dz \approx p'(0,V)/k_{TF}$ as $Q(V) \approx tp_o(\frac{R_s(0\ V)}{R_s(V)} - 1)$.

Here, the thickness is 3.3 nm, the initial hole density is estimated from previous Hall effect measurements as 1.06 x $10^{22}$ $cm^{-3}$,[3] and we take $R_s$(0 V) and $R_s$(V) from Fig. 3b.

The calculated accumulated sheet density at 147 °C is plotted in Fig. 4c. At this temperature, the IL is in the liquid phase and the EDL capacitance should be as expected from the literature on room temperature ILs. It can be seen that the sheet density is ~1-2 x $10^{14}$ $cm^{-2}$ for $V_G$ up to -2.5 V, in the range expected for IL EDLs.[23] The volume of the anion used here, as calculated using Chem3D, is 200.55 $Å^3$, corresponding to a sheet density of 2-4 x $10^{14}$ $cm^{-2}$ for one complete monolayer of the anion, assuming anion rigidity. Thus, the calculated sheet density in Fig. 4c also agrees with the expected density from considerations of the anion volume. As a function of gate voltage, $Q$ is rather linear, indicating that the capacitance of the EDL is nearly constant in this temperature and voltage regime. From the slope of the $Q$-$V$ curve we find that the capacitance is ~12 $\mu F/cm^2$, in good agreement with the EDL capacitance of similar IL molecules measured on reference electrodes.[5] Note that determining the EDL capacitance in this way presumes that the IL-gate electrode interface has negligible voltage drop, which is a valid assumption if the IL-gate contact area (and hence capacitance) is significantly larger than the IL-SNO contact area. A non-zero voltage drop at the IL-gate interface increases uncertainty in the extracted IL-SNO capacitance, but likely not by more than a factor of 2, corresponding to equal voltage drop across both interfaces. In either case, EDL capacitance up to 25 $\mu F/cm^2$ is still within the range observed for other IL-gated systems.[23] The agreement between calculated and literature values of the accumulated sheet charge density and EDL capacitance strengthens the above model for electrostatic gating of SNO thin films. It also strengthens the assertion that electrochemical effects are negligible in our devices because it is known that electrochemical Faradaic reactions do not result in mobile charge accumulation.[8] Thus, we would not expect the above model to simulate well a resistance modulation due to chemical reactions. Note that we only model the accumulated charge at a fixed high temperature in which the IL is in the liquid phase. This is because the temperature-dependence of the accumulation layer SNO resistance and the EDL capacitance are not well understood, particularly for ILs in the solid phase.

In conclusion, we have investigated room temperature electrostatic gating of ultrathin $SmNiO_3$ films using an ionic liquid. Proper usage of the ionic liquid is crucial to achieve reproducible gating, and water incorporation into the ionic liquid may be responsible for electrochemical etching of the $SmNiO_3$ film. With increasing negative gate voltage, the $SmNiO_3$ sheet resistance decreases in both insulating and metallic phases, up to nearly 25% reduction at room temperature, and the metal-insulator transition temperature concurrently decreases. The electrostatically accumulated sheet charge density and electric double layer capacitance was modeled based on an exponential decay of the hole density away from the $SmNiO_3$ surface, and good agreement with expected values from literature was found.

**Acknowledgements**

The authors acknowledge the ARO MURI (W911-NF-09-1-0398), and AFOSR (FA9550-12-1-0189) for financial support. This work was performed in part at the Center for Nanoscale Systems at Harvard University, which is supported under NSF award ECS-0335765. The authors thank P. Ehrhardt for help with the chlorine irradiation and Prof. X. D. Wang for usage of STEM facilities.

**Figure captions**

**Figure 1:** **(a)** Cross sectional STEM image of SNO film deposited on LAO as used in this study. Pt layer is used as protective cap during STEM sample preparation. Zone axis of LAO is labeled. Inset: High-resolution zoom of interface. Arrow denotes interface. **(b)** Rendering of three-terminal SNO device. IL overlaps with SNO bar and gate electrode. Gate voltage is applied with source terminal grounded. **(c)** Schematic of cation (left) and anion (right) molecules in the IL used in this work.

**Figure 2:** $\rho$-$T$ curves for SNO films with high oxygen content (~370 mTorr growth, black curve), low oxygen content (~285 mTorr, blue curve), and high oxygen content after irradiation with $Cl^{2+}$ ions (red curve).

**Figure 3:** $R_s$-$T$ data for SNO three-terminal devices gated **(a)** with the IL used as-received and **(b)** with the IL used after baking in $N_2$ flow and measuring in $N_2$. Gate voltages are as indicated in the legends.

**Figure 4:** **(a)** Percent change in sheet resistance $(R_s(V) - R_s(0))/R_s(0)$ as a function of temperature and gate voltage. **(b)** $T_{MI}$ as a function of gate voltage. **(c)** Calculated accumulated sheet charge density as a function of gate voltage from resistance data at 147 °C. EDL capacitance extracted from the slope of a linear fit to the $Q$-$V$ data. Inset: Illustration of assumed hole density profile as a function of distance away from the IL-SNO interface.

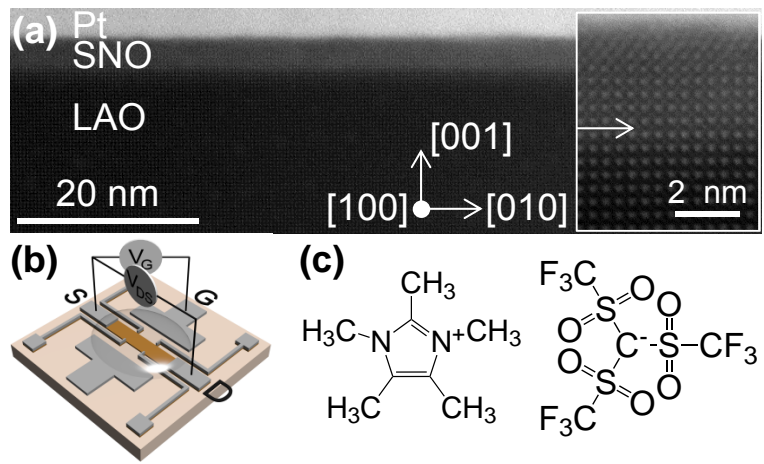

Figure 1

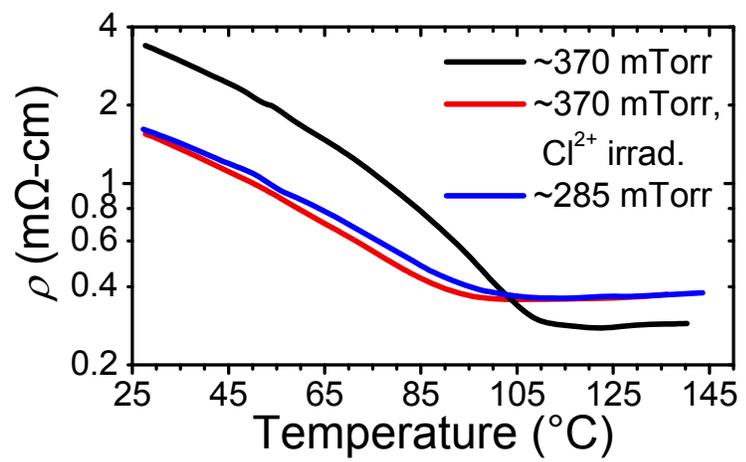

Figure 2

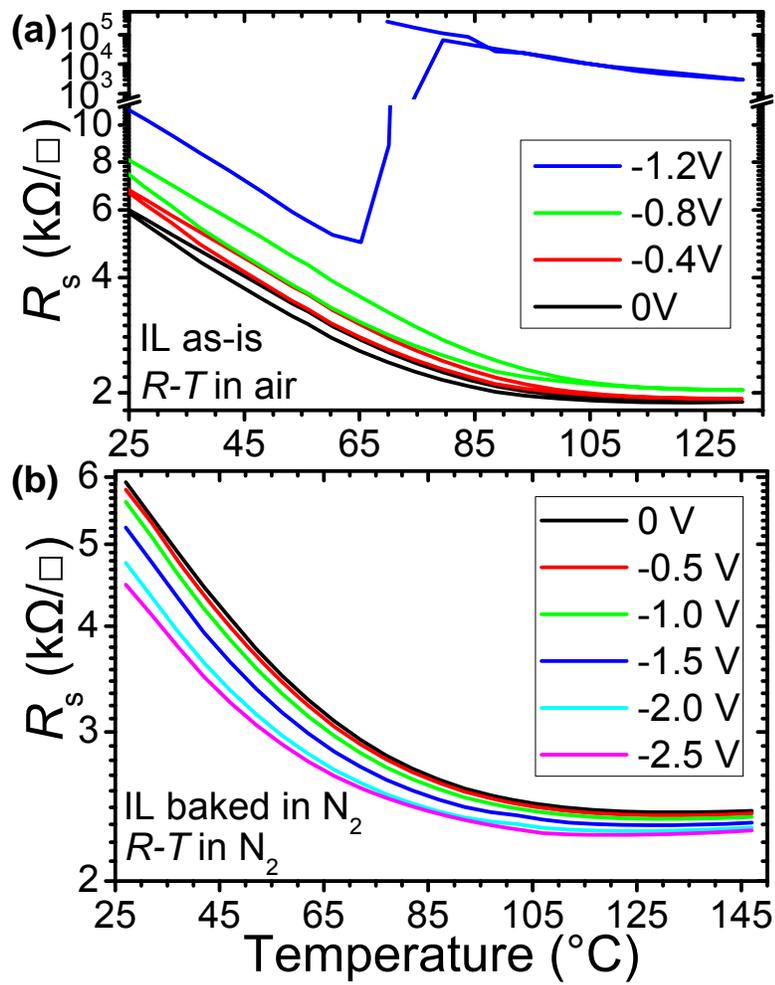

Figure 3

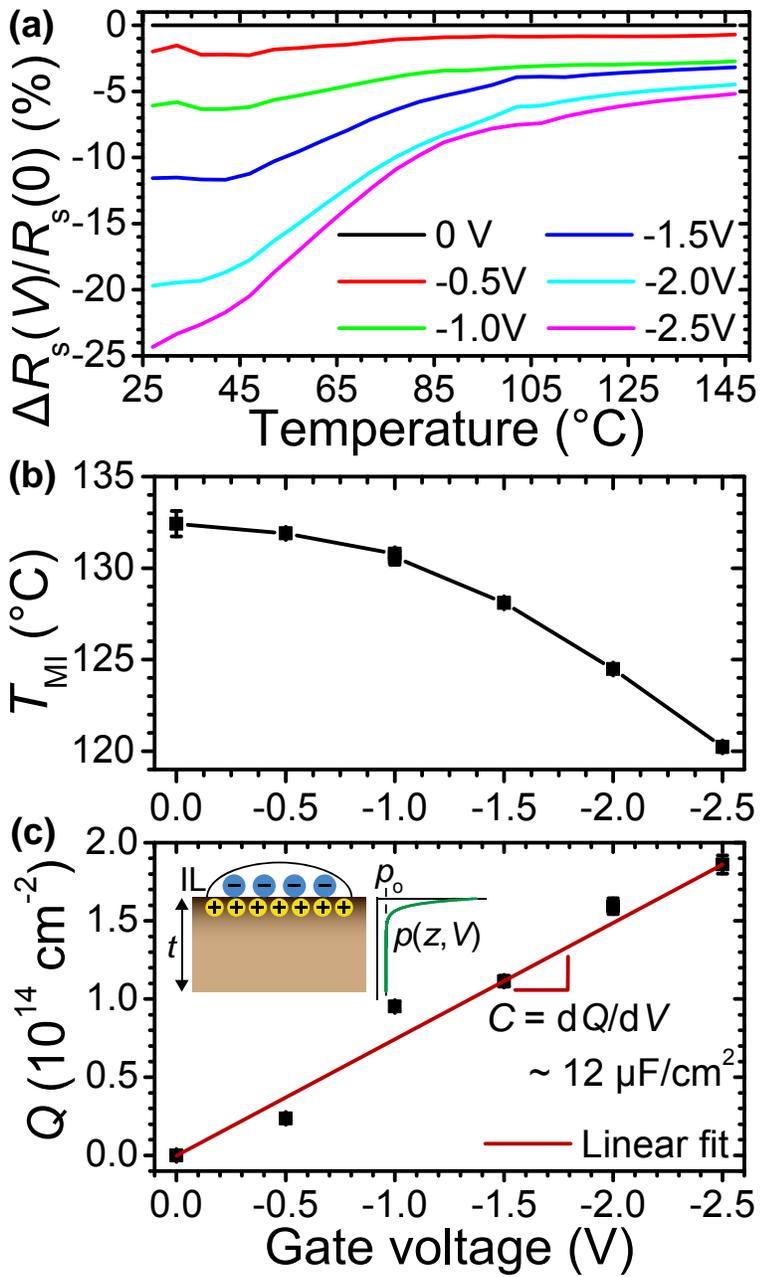

Figure 4